\documentclass{article}
\usepackage{amssymb}

\usepackage{fleqn}
\usepackage{epsfig}
\usepackage{graphicx}
\usepackage{amsmath}

\input{tcilatex}
\def\beq{\begin{equation}}
\def\eeq{\end{equation}}
\def\bea{\begin{eqnarray}}
\def\eea{\end{eqnarray}}
\def\bq{\begin{quote}}
\def\eq{\end{quote}}

\def\gappeq{\mathrel{\rlap {\raise.5ex\hbox{$>$}}
{\lower.5ex\hbox{$\sim$}}}}
\def\lappeq{\mathrel{\rlap{\raise.5ex\hbox{$<$}}
{\lower.5ex\hbox{$\sim$}}}}
\def\Toprel#1\over#2{\mathrel{\mathop{#2}\limits^{#1}}}

\begin{document}

\pagestyle{empty} 
\begin{flushright}
{CERN-TH/2001-035}\\
hep-ph/0102138\\
\end{flushright}
\vspace*{15mm}

\begin{center}
\textbf{RESUMMED COEFFICIENT FUNCTION FOR THE SHAPE FUNCTION } \\[0pt]
\vspace*{1cm} \textbf{U. Aglietti}$^{\ast )}$ \\[0pt]
\vspace{0.3cm} Theoretical Physics Division, CERN\\[1pt]
CH - 1211 Geneva 23 \\[3pt]
$~~~$ \\[0pt]
\vspace*{2cm} \textbf{Abstract} \\[0pt]
\end{center}

We present a leading evaluation of the resummed coefficient function for the
shape function. It is also shown that the coefficient function is
short-distance-dominated. Our results allow relating the shape function
computed on the lattice to the physical QCD distributions. \vspace*{7cm}
\noindent %\rule[.1in]{16.5cm}{.002in}

\noindent $^{\ast )}$ On leave of absence from Dipartimento di Fisica,
Universita' di Roma I, Piazzale Aldo Moro 2, 00185 Roma, Italy. E-mail
address: ugo.aglietti@cern.ch. \vspace*{0.5cm}

\begin{flushleft} CERN-TH/2001-035 \\[0pt]
February 2001
\end{flushleft}
\vfill\eject
%\pagestyle{empty}
%\clearpage\mbox{}\clearpage

\setcounter{page}{1} \pagestyle{plain}

\section{Introduction \noindent}

In this note we present a leading evaluation of the coefficient function of
the shape function \cite{nostri}--\cite{giuliame}. The latter is also called
structure function of the heavy flavours. The coefficient function allows
relating the shape function --- computed with a non-perturbative technique
such as lattice QCD --- to distributions in semi-inclusive heavy-flavour
decays. We consider in particular the processes 
\begin{equation}
B\rightarrow X_{s}+\gamma  \label{rare}
\end{equation}
and 
\begin{equation}
B\rightarrow X_{u}+l+\nu  \label{semi}
\end{equation}
in the hard limit 
\begin{equation}
Q\gg \Lambda .  \label{hardlim}
\end{equation}
The quantity $Q$ is the hard scale of these time-like processes: $Q\equiv 2%
\mathcal{E}$ , where $\mathcal{E}$ is the hadronic energy of the state $X$
and $\Lambda $ is the QCD scale\footnote{%
For the rare decay (\ref{rare}), one can actually set $Q=m_{B}.$}. For a $B$%
-meson at rest, $v=\left( 1;0,0,0\right) ,$ and the jet $X$ flying along the
minus direction ($-z$ axis), the shape function is defined as 
\begin{equation}
\varphi \left( k_{+}\right) \equiv \langle B\left( v\right) |h_{v}^{\dagger
}\delta \left( k_{+}-iD_{+}\right) h_{v}|B\left( v\right) \rangle .
\label{shape}
\end{equation}
This represents the probability that the $b$-quark in the $B$-meson has
momenta 
\begin{equation}
p_{b}=m_{B}v+k^{\prime }  \label{perme}
\end{equation}
with any transverse and minus components and with given plus component 
\begin{equation}
k_{+}^{\prime }=k_{+}.
\end{equation}
The static field $h_{v}\left( x\right) $ is related to the Dirac field of
the beauty quark $b\left( x\right) $ by 
\begin{equation}
b\left( x\right) =e^{-im_{B}v\cdot x}h_{v}\left( x\right) +O\left( \frac{1}{%
m_{B}}\right) .
\end{equation}
With the shape function, the decays (\ref{rare}) and (\ref{semi}) are
related to their respective quark-level processes 
\begin{equation}
b\rightarrow \widehat{X}_{s}+\gamma
\end{equation}
and 
\begin{equation}
b\rightarrow \widehat{X}_{u}+l+\nu ,
\end{equation}
where the $b$-quark has the momentum (\ref{perme}) with the distribution (%
\ref{shape}). It can be shown that the invariant mass $m$ of the state $%
\widehat{X}$ is related to the virtuality of the $b$-quark by 
\begin{equation}
k_{+}=-\frac{m^{2}}{Q}.
\end{equation}
The shape function is a non-perturbative distribution --- analogous to
parton distribution functions --- and it describes the slice of the
semi-inclusive region\footnote{%
This region is also called threshold region, large-$x$ region,
radiation-inhibited region and Sudakov region.} in which 
\begin{equation*}
m^{2}\sim \Lambda \,Q,
\end{equation*}
i.e. 
\begin{equation}
|k_{+}|\sim \Lambda ,\qquad \mathrm{or\quad }\quad z\sim 1-\frac{\Lambda }{Q}%
,  \label{cinshape}
\end{equation}
where 
\begin{equation}
z\equiv 1-\frac{m^{2}}{Q^{2}}.
\end{equation}
Because of ultraviolet divergences affecting its matrix elements, the shape
function has a dependence on the ultraviolet cut-off or renormalization
point $\mu :$%
\begin{equation}
\varphi \left( k_{+}\right) =\varphi \left( k_{+};\mu \right) ,
\end{equation}
and is related to a physical QCD distribution by means of a coefficient
function by (cf. eq.\thinspace (\ref{ba})) 
\begin{equation}
\mathbf{\varphi }\left( k_{+};Q\right) =\int dk_{+}^{\prime }\,C\left(
k_{+}-k_{+}^{\prime };Q,\mu \right) \,\varphi \left( k_{+}^{\prime };\mu
\right) .  \label{definente}
\end{equation}
The QCD distribution does not depend on $\mu :$ 
\begin{equation}
\frac{d}{d\mu }\mathbf{\varphi }\left( k_{+};Q\right) =0,
\end{equation}
while the shape function does not depend on $Q.$ As a consequence, the
coefficient function depends on both $Q$ and $\mu .$

The shape function can be computed with a non-perturbative technique or
extracted from experimental data. If it is computed inside a field-theory
model --- such as lattice QCD --- its expression will exhibit a $\mu $%
-dependence that cancels against that of the coefficient function. If it is
instead computed inside a phenomenological model --- such as a quark model
--- the situation is less transparent. The $\mu $-independence is not
``automatic'' and one has to specify the value of \ $\mu $ appropriate for
the model. Some care is needed also in extracting the shape function from
the experimental data, in order to avoid double counting of perturbative
corrections. A factorization scheme must be defined and the coefficient
functions for the various processes all have to be computed in the same
scheme\footnote{%
The situation is analogous to usual hard processes, where various
factorization schemes for the parton distribution functions are defined:
DIS, $\overline{\mathrm{MS}},$ etc.}. In particular, if a branching
MonteCarlo is used for the analysis, the perturbative corrections generated
by the program must be subtracted.

The coefficient function is obtained by evaluating in leading approximation
both $\mathbf{\varphi }$ and $\varphi $ and inserting their expression in
eq.\thinspace (\ref{definente}). Since the coefficient function is expected
to be a short-distance quantity, we compute the QCD distribution and the
shape function in PT for an on-shell $b$-quark ($k^{\prime }=0$). This
expectation will be verified \textit{a posteriori. }

\section{The QCD distribution}

The (perturbative) long-distance effects occurring in (\ref{rare}) and (\ref
{semi}) can be factorized in the function 
\begin{equation}
\mathbf{f}\left( z\right) =\delta \left( 1-z-0\right) -A_{1}\alpha
_{S}\left( \frac{\log \left[ 1-z\right] }{1-z}\right) _{+},
\label{verificata}
\end{equation}
where 
\begin{equation}
A_{1}=\frac{C_{F}}{\pi }
\end{equation}
and $C_{F}=\left( N_{c}^{2}-1\right) /\left( 2N_{c}\right) =4/3.$ The
plus-distribution is defined as usual as 
\begin{equation}
P\left( z\right) _{+}\equiv P\left( z\right) \,-\,\delta \left( 1-z-0\right)
\int_{0}^{1}dy\,P\left( y\right) .
\end{equation}
The integrated or cumulative distribution is defined as 
\begin{equation}
\mathbf{F}\left( z\right) \equiv \int_{z}^{1}dz^{\prime }\,\mathbf{f}\left(
z^{\prime }\right) .
\end{equation}
Inserting expression (\ref{verificata}) in this, one obtains the well-known
double logarithm: 
\begin{equation}
\mathbf{F}\left( z\right) =1-\frac{A_{1}\alpha _{S}}{2}\log ^{2}\left(
1-z\right) .
\end{equation}
The cumulative distribution satisfies the normalization condition $\mathbf{F}%
\left( 0\right) =1.$ Multiple soft-gluon emission exponentiates the one-loop
distribution, so that 
\begin{equation}
\mathbf{F}\left( z\right) =\exp \left[ -\frac{A_{1}\alpha _{S}}{2}\log
^{2}\left( 1-z\right) \right] .  \label{frozen}
\end{equation}
For further improvement, it is convenient to write the function $\mathbf{f}%
\left( z\right) $ in an ``unintegrated'' form, as 
\begin{equation}
\mathbf{f}\left( z\right) =\delta \left( 1-z\right) +A_{1}\alpha
_{S}\int_{0}^{1}\frac{d\epsilon }{\epsilon }\int_{0}^{1}\frac{dt}{t}\left[
\delta \left( 1-z-\epsilon t\right) -\delta \left( 1-z\right) \right] ,
\label{bo}
\end{equation}
where we have defined the unitary energy and angular variables 
\begin{equation}
\epsilon \equiv \frac{E}{Q}\qquad \mathrm{and}\qquad t\equiv \frac{1-\cos
\theta }{2}.
\end{equation}
The quantity $E$ is two times the energy of the soft gluon, $E=2E_{g},$ and $%
\theta $ is the emission angle. Leading logarithmic corrections are included
replacing the bare coupling with the running coupling evaluated at the gluon
transverse momentum squared \cite{ven}: 
\begin{equation}
\alpha _{S}\rightarrow \alpha _{S}\left( l_{\perp }^{2}\right) ,
\end{equation}
with 
\begin{equation}
l_{\perp }^{2}\simeq E_{g}^{2}\,\theta ^{2}\simeq Q^{2}\epsilon ^{2}t.
\end{equation}
The QCD semi-inclusive form factor then reads: 
\begin{equation}
\mathbf{f}\left( z\right) =\delta \left( 1-z\right) +\int_{0}^{1}\frac{%
d\epsilon }{\epsilon }\int_{0}^{1}\frac{dt}{t}A_{1}\alpha _{S}\left(
Q^{2}\epsilon ^{2}t\right) \left[ \delta \left( 1-z-\epsilon t\right)
-\delta \left( 1-z\right) \right] .  \label{frozen2}
\end{equation}
Performing the integrations, one obtains: 
\begin{equation}
\mathbf{f}\left( z\right) =\delta \left( 1-z\right) +\frac{A_{1}}{\beta _{0}}%
\left\{ \left[ \frac{\log \log \xi \left( 1-z\right) }{1-z}\right] _{+}-%
\left[ \frac{\log \log \xi \left( 1-z\right) ^{2}}{1-z}\right] _{+}\right\} ,
\label{running2}
\end{equation}
where $\beta _{0}\equiv \left( 11/3\,N_{c}-2/3\,n_{f}\right) /\left( 4\pi
\right) $ and $\xi $ is the square of the ratio of the hard scale to the QCD
scale: 
\begin{equation}
\xi \equiv \frac{Q^{2}}{\Lambda ^{2}}.
\end{equation}
Equation\thinspace (\ref{running2}) replaces the frozen-coupling result (\ref
{verificata}) and reduces to it in the limit $\xi \rightarrow \infty $ with $%
z$ fixed. The integrated distribution reads 
\begin{equation}
\mathbf{F}\left( z\right) =1-\frac{A_{1}}{2\beta _{0}}\left[ \log \xi \log
\log \xi -2\log \xi \left( 1-z\right) \log \log \xi \left( 1-z\right) +\log
\xi \left( 1-z\right) ^{2}\log \log \xi \left( 1-z\right) ^{2}\right] .
\end{equation}
As in the frozen-coupling case, higher-order corrections exponentiate the
one-loop result, so that 
\begin{equation}
\mathbf{F}\left( z\right) =\exp \left\{ -\frac{A_{1}}{2\beta _{0}}\left[
\log \xi \log \log \xi -2\log \xi \left( 1-z\right) \log \log \xi \left(
1-z\right) +\log \xi \left( 1-z\right) ^{2}\log \log \xi \left( 1-z\right)
^{2}\right] \right\} .  \label{running}
\end{equation}
Equation\thinspace (\ref{running}) replaces the frozen-coupling result (\ref
{frozen}).

The above distributions are ``physical'' and therefore should be real for
any value of $z$ in the range 
\begin{equation}
0\leq z\leq 1.
\end{equation}
In practice, the distributions (\ref{running2}) and (\ref{running}) are real
only if the range of $z$ is restricted to 
\begin{equation}
z<1-\frac{\Lambda }{Q}\qquad \mathrm{or\qquad }m^{2}>\Lambda \,Q.
\end{equation}
This restriction is absent in the frozen coupling case and originates from
the infrared pole in the running coupling, which diverges when the
transverse gluon momentum becomes as small as the QCD scale. The resummed
distribution is therefore reliable as long as one stays away from the
infrared pole, which implies the limitation 
\begin{equation}
m^{2}\gg \Lambda \,Q.  \label{nonso}
\end{equation}
This condition makes the kinematic region (\ref{cinshape}) unaccessible to
the QCD distribution and forces the introduction of a non-perturbative
component, namely the shape function. The physical origin of the restriction
(\ref{nonso}) is easily understood with the following qualitative
considerations. The jet mass and the gluon transverse momentum are given by 
\begin{eqnarray}
\qquad \qquad \qquad \qquad \qquad \qquad \qquad \qquad \qquad \frac{m^{2}}{Q%
} &\sim &E\,\theta ^{2},  \notag \\
l_{\perp }\,\, &\sim &E\,\theta .  \label{nuove}
\end{eqnarray}
The smallest transverse momentum for fixed invariant mass is obtained for $%
\theta \sim 1$ and is 
\begin{equation}
l_{\perp \min }\sim \frac{m^{2}}{Q}.  \label{ultima}
\end{equation}
In region (\ref{cinshape}) this momentum is of the order of the QCD scale,
where the coupling leaves the perturbative phase. The conclusion is that
large-angle soft-gluon emission signals non-perturbative effects in the
region (\ref{cinshape}).

In general, resummation of multiple soft-gluon emission is performed in
Mellin space. We therefore consider the moments of the form factor: 
\begin{eqnarray}
\qquad \qquad \qquad \qquad \qquad \qquad \qquad \qquad \mathbf{f}_{N}
&\equiv &\int_{0}^{1}dz\,z^{N}\,\mathbf{f}\left( z\right)  \notag \\
&=&1+\Delta \mathbf{f}_{N}.
\end{eqnarray}
Exponentiation of the ``effective''\footnote{%
We call it ``effective'' because the insertion of the running coupling in
the time-like region already includes some multiple gluon-emission effects.}
one-gluon distribution takes place, so that 
\begin{equation}
\mathbf{f}_{N}=e^{\Delta \mathbf{f}_{N}}.
\end{equation}
Using the large-$N$ approximation \cite{cattren} 
\begin{equation}
\left( 1-y\right) ^{N}-1\simeq -\theta \left( y-1/n\right) ,  \label{approx}
\end{equation}
where $n\equiv N/N_{0}$ with $N_{0}\equiv e^{-\gamma _{E}}\simeq 0.56$ and $%
\gamma _{E}\simeq 0.58$ is the Euler constant, the moments of eq. (\ref
{verificata}) or (\ref{bo}) read (frozen-coupling case, $\beta
_{0}\rightarrow 0$) 
\begin{equation}
\mathbf{f}_{n}=\exp \left[ -\frac{A_{1}\alpha _{S}}{2}\log ^{2}n\right] .
\label{fixed3}
\end{equation}
In the running coupling case, the Mellin transform of eq. (\ref{frozen2}) or
(\ref{running2}) is \cite{civuole, kodtren, cattren}: 
\begin{equation}
\mathbf{f}_{n}=\exp \left\{ -\frac{A_{1}}{2\beta _{0}}\left[ \log \frac{\xi 
}{n^{2}}\log \log \frac{\xi }{n^{2}}-2\log \frac{\xi }{n}\log \log \frac{\xi 
}{n}+\log \xi \log \log \xi \right] \right\} .  \label{baba}
\end{equation}
Note that $\mathbf{f}_{n}=1$ for $n=1.$ The distribution (\ref{baba}) is
usually written as 
\begin{equation}
\mathbf{f}_{n}=\exp \left[ l\,\,g_{1}\left( \beta _{0}\alpha _{S}l\right) 
\right] ,
\end{equation}
where $l\equiv \log n,$ $\alpha _{S}\equiv \alpha _{S}\left( Q\right) $ and 
\begin{equation}
\,g_{1}\left( w\right) \equiv -\frac{A_{1}}{2\beta _{0}}\,\frac{1}{w}\left[
\left( 1-2w\right) \log \left( 1-2w\right) -2\left( 1-w\right) \log \left(
1-w\right) \right] .
\end{equation}
Expanding this function to lowest order in $w,$ one obtains $g_{1}\left(
w\right) =-A_{1}w/(2\beta _{0})+O\left( w^{2}\right) $ and recovers the
fixed coupling result.

As we clearly see, the inverse transform from $\mathbf{f}_{n}$ to $\mathbf{F}%
\left( z\right) $ is simply computed with the replacement \cite{mangano} 
\begin{equation}
n\rightarrow \frac{1}{1-z}.  \label{stima}
\end{equation}
The limitation to the resummed perturbative result found before
(eq.\thinspace (\ref{nonso})) reads, in $N$-space: 
\begin{equation}
n\ll \frac{Q}{\Lambda }.  \label{limitati}
\end{equation}

\section{The shape function}

Let us now consider the quantity in the effective theory related to $\mathbf{%
f}\left( z\right) $, namely the shape function $\varphi \left( k_{+}\right)
. $ For an on-shell $b$-quark, the latter is given by 
\begin{equation}
\varphi \left( k_{+}\right) =\delta \left( k_{+}\right) +A_{1}\alpha _{S}\,%
\frac{\theta \left( 0;k_{+};-\mu \right) }{-k_{+}}\log \frac{\mu }{-k_{+}}%
-A_{1}\alpha _{S}\,\delta \left( k_{+}\right) \int_{-\mu }^{0}\frac{dl_{+}}{%
-l_{+}}\log \frac{\mu }{-l_{+}},
\end{equation}
where $\mu \equiv 2\Lambda _{S}$ and we defined $\theta \left(
a_{1};a_{2};\cdots a_{n}\right) \equiv \theta \left( a_{1}-a_{2}\right)
\theta \left( a_{2}-a_{3}\right) \cdots \theta \left( a_{n-1}-a_{n}\right) .$
The regularization used \cite{giuliame} imposes a cutoff on the spatial loop
momenta and not on the energies: 
\begin{equation}
|\overrightarrow{l}|<\Lambda _{S},\qquad -\infty <l_{0}<+\infty ;
\end{equation}
it is qualitatively similar to lattice regularization. Proceeding in a
similar way, one can write 
\begin{equation}
\varphi \left( k_{+}\right) =\delta \left( k_{+}\right) +\int_{0}^{\mu }%
\frac{dE}{E}\int_{0}^{1}\frac{dt}{t}A_{1}\alpha _{S}\left( E^{2}t\right) %
\left[ \delta \left( k_{+}+Et\right) -\delta \left( k_{+}\right) \right] .
\label{ancoralei}
\end{equation}
We explicitly see that the hard scale $Q$ does not appear in $f\left(
k_{+}\right) $, as it should. Since $k_{+}=-Q\left( 1-z\right) ,$ the shape
function in the ``QCD variable'' $z$ reads 
\begin{equation}
f\left( z\right) =\delta \left( 1-z\right) +\int_{0}^{\eta }\frac{d\epsilon 
}{\epsilon }\int_{0}^{1}\frac{dt}{t}A_{1}\alpha _{S}\left( Q^{2}\epsilon
^{2}t\right) \left[ \delta \left( 1-z-\epsilon t\right) -\delta \left(
1-z\right) \right] ,
\end{equation}
where we have defined an adimensional shape function as 
\begin{equation}
f\left( z\right) \equiv Q\,\varphi \left( k_{+}\right) .
\end{equation}
The quantity $\eta $ is the ratio of the UV cutoff of the effective theory
to the hard scale, 
\begin{equation}
\eta \equiv \frac{\mu }{Q}<1,
\end{equation}
because the shape function is defined in a low-energy effective theory. To
avoid substantial finite cut-off effects, it must also be assumed that 
\begin{equation}
\mu \gg \Lambda .  \label{ovvia}
\end{equation}
Taking the Mellin moments and exponentiating the one-loop distribution as in
the QCD case, one obtains\footnote{%
Do not confuse \textit{these} moments of the shape function $f_{N}$ with the
moments considered by other authors, $\int dk_{+}\,k_{+}^{N}\,\varphi \left(
k_{+}\right) \sim \int dz\,(1-z)^{N}\,f\left( z\right) .$} 
\begin{equation}
f_{N}=e^{\Delta f_{N}}.
\end{equation}
A straightforward computation gives 
\begin{equation}
\log f_{n}=-\frac{A_{1}}{2\beta _{0}}\theta \left( n-1/\eta \right) \left[
\log \frac{\xi }{n^{2}}\log \log \frac{\xi }{n^{2}}-2\log \frac{\eta \xi }{n}%
\log \log \frac{\eta \xi }{n}+\log \eta ^{2}\xi \log \log \eta ^{2}\xi 
\right] .
\end{equation}
As in the case of the QCD distribution, $f_{n}=1$ for $n=1.$ The $N$-moments
of the shape function are reliably computed in PT if one restricts $n$ as in
eq.\thinspace (\ref{limitati}). We therefore have here the same limitation
of the QCD distribution, as expected on physical ground.

\section{The coefficient function}

We now introduce the coefficient function $C\left( z\right) ,$ relating the
shape function to the QCD semi-inclusive form factor, as 
\begin{equation}
C\left( z\right) \equiv \delta \left( 1-z\right) +\Delta C\left( z\right) ,
\end{equation}
where 
\begin{equation}
\Delta C\left( z\right) \equiv \mathbf{f}\left( z\right) -\,f\left( z\right)
.  \label{diffe}
\end{equation}
Inserting the expressions for the two distributions, we obtain 
\begin{equation}
\Delta C\left( z\right) =\int_{\eta }^{1}\frac{d\epsilon }{\epsilon }%
\int_{0}^{1}\frac{dt}{t}A_{1}\alpha _{S}\left( Q^{2}\epsilon ^{2}t\right) %
\left[ \delta \left( 1-z-\epsilon t\right) -\delta \left( 1-z\right) \right]
.  \label{ancora}
\end{equation}
The expression for the coefficient function is similar to the one for the
QCD form factor, the only difference being a lower cut-off on the gluon
energies in this case. Note that there is no angular cutoff. The QCD
distribution and the shape function are related by a convolution in momentum
space, 
\begin{equation}
\mathbf{f}\left( z\right) =\int_{0}^{1}\int_{0}^{1}dz^{\prime }dz^{\prime
\prime }\delta \left( z-z^{\prime }z^{\prime \prime }\right) \,C\left(
z^{\prime }\right) \,f\left( z^{\prime \prime }\right) .  \label{fondamento2}
\end{equation}
Taking the moments on both sides, one diagonalizes the convolution, so that 
\begin{equation}
\mathbf{f}_{n}=\,C_{n}\,f_{n}.  \label{fondamento}
\end{equation}
The usual exponentiation reads: 
\begin{equation}
C_{n}=e^{\Delta C_{n}}.
\end{equation}
The coefficient function can be computed from the difference in
eq.\thinspace (\ref{diffe}) or directly from the integral representation in
eq.\thinspace (\ref{ancora}) in the following way. The integration over the
emission angle $t$ gives 
\begin{equation}
\Delta C_{N}=\int_{0}^{1}\frac{dy}{y}\left[ \left( 1-y\right) ^{N}-1\right]
\int_{\max \left( \eta ,\,y\right) }^{1}\frac{d\epsilon }{\epsilon }%
A_{1}\alpha _{S}\left( Q^{2}\epsilon y\right) ,
\end{equation}
where $y\equiv 1-z.$ The integration over $y$ is done using the previous
approximation (\ref{approx}) and one finally obtains: 
\begin{eqnarray}
\log C_{n} &=&-\frac{A_{1}}{2\beta _{0}}\left\{ \theta \left( 1/\eta
-n\right) \left[ \log \frac{\xi }{n^{2}}\log \log \frac{\xi }{n^{2}}-2\log 
\frac{\xi }{n}\log \log \frac{\xi }{n}+\log \xi \log \log \xi \right]
\right. +  \label{lei} \\
&&\left. +\theta \left( n-1/\eta \right) \left[ 2\log \frac{\eta \xi }{n}%
\log \log \frac{\eta \xi }{n}\,\,\,-2\log \frac{\xi }{n}\log \log \frac{\xi 
}{n}+\log \xi \log \log \xi -\log \eta ^{2}\xi \log \log \eta ^{2}\xi \right]
\right\} .  \notag
\end{eqnarray}
The QCD function $\mathbf{f}_{n}$ is equal to the coefficient of $\theta
\left( 1/\eta -n\right) $ in the r.h.s of eq.\thinspace (\ref{lei}).
\noindent Collecting the terms in common to the two $\theta $-functions,
eq.\thinspace (\ref{lei})\thinspace can also be written as: 
\begin{eqnarray}
\log C_{n} &=&\left. -\frac{A_{1}}{2\beta _{0}}\right\{ \theta \left( 1/\eta
-n\right) \log \frac{\xi }{n^{2}}\log \log \frac{\xi }{n^{2}}+\log \xi \log
\log \xi -2\log \frac{\xi }{n}\log \log \frac{\xi }{n}+  \notag \\
&&\left. \qquad \,\,\,\,+\theta \left( n-1/\eta \right) \,\left[ 2\,\log \,%
\frac{\eta \xi }{n}\,\log \log \,\frac{\eta \xi }{n}-\log \eta ^{2}\xi \log
\log \eta ^{2}\xi \right] \right\} .  \label{lei2}
\end{eqnarray}
Let us comment on the result, represented by eq.\thinspace (\ref{lei}) or by
eq.\thinspace (\ref{lei2}). It is immediate to check that $C_{n}=1$ when $%
\eta =1$ or when $n=1,$ as it should. Furthermore, $C_{n}\rightarrow 1$ in
the limit $Q\rightarrow \infty $ with $n$ and $\mu $ fixed.

\begin{figure}[tbh]
\centerline{\mbox{\epsfxsize=6cm\epsffile{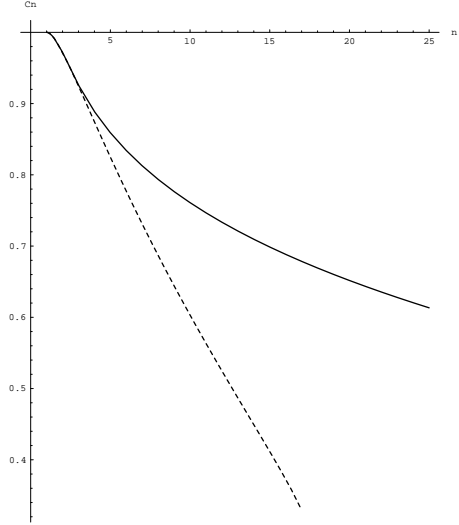}}}
\caption{ Plot of the resummed coefficient function $C_{n}$ (solid line) and
the QCD semi-inclusive form factor $\mathbf{f}_{n}$ (dashed line) for the
choice of the parameters discussed in the text.}
\label{figure}
\end{figure}
The main point is that the coefficient function is perturbatively computable
as long as 
\begin{equation*}
n\,\ll \,\frac{Q\,\mu }{\Lambda ^{2}}.
\end{equation*}
Note that this critical value of $n$ is much larger than the one for the QCD
form factor or the shape function given in eq.\thinspace (\ref{limitati}),
because of eq.\thinspace (\ref{ovvia}). The equivalent limitation in
momentum space for $C\left( z\right) $ is 
\begin{equation}
m^{2}\,\gg \,\frac{Q\,\Lambda ^{2}}{\mu }.  \label{limit}
\end{equation}
The occurrence of non-perturbative effects outside the region indicated by
eq.\thinspace (\ref{limit}) can be understood in a simple way. The problem
is to find the smallest gluon transverse momentum for a given invariant
mass, eqs.\thinspace (\ref{nuove}), with the additional constraint $E\gtrsim
\mu .$ The minimum occurs for the smallest gluon energy, $E\sim \mu ,$ where
one has 
\begin{equation}
\theta \,\sim \,\sqrt{\frac{m^{2}}{\mu \,Q}}\,\sim \,\sqrt{\frac{\Lambda }{%
\mu }}\,\ll \,1
\end{equation}
and 
\begin{equation}
l_{\perp \min }\,\sim \,\sqrt{\frac{m^{2}}{Q}\mu }\,\sim \,\sqrt{\Lambda
\,\mu }\gg \,\Lambda ,  \label{ultimab}
\end{equation}
in which conditions (\ref{cinshape}) and (\ref{ovvia}) have been used. A
comparison between eqs.\thinspace (\ref{ultima}) and (\ref{ultimab}) shows
that transverse momenta are substantially larger in the coefficient function
than in the QCD distribution; this allows a perturbative treatment of the
former. This is because the infrared cut-off on the gluon energies in $%
C\left( z\right) $ has the effect of lowering the largest emission angle,
indirectly increasing the minimum gluon transverse momentum.

In single logarithmic problems, the loop expressions for the coefficient
functions usually have phase-space restrictions of the form 
\begin{equation}
l_{\perp }\,\gtrsim \,\mu \,\gg \,\Lambda .
\end{equation}
Therefore, factorization with the shape function involves, as the relevant
dynamical scale, a softer scale with respect to single logarithmic problems,
because 
\begin{equation}
\mu \,\gg \,\sqrt{\Lambda \,\mu }.
\end{equation}
Non-perturbative corrections to the factorized form (\ref{fondamento2}) or (%
\ref{fondamento})\ are expected to be of the size 
\begin{equation}
\frac{\Lambda ^{2}}{l_{\perp \min }^{2}},  \label{stima2}
\end{equation}
which means of order 
\begin{equation}
\frac{\Lambda ^{2}}{|k_{+}|\,\mu }\sim \frac{\Lambda }{\mu }.
\end{equation}
One inverse power of the factorization scale $\mu $ is involved. The
corrections are therefore much larger than in single logarithmic problems,
where the estimate (\ref{stima2}) translates to 
\begin{equation}
\frac{\Lambda ^{2}}{\mu ^{2}},
\end{equation}
involving two inverse powers of the factorization scale.

The previous findings can be summarized by considering eq. (\ref{fondamento}%
) in the various regions of the moment index $n,$ corresponding to different
dynamical regimes:

\begin{enumerate}
\item  In the region\footnote{%
The region $n\sim 1$ is trivial as there are no large infrared logarithms
and fixed order perturbation theory is sufficient.} 
\begin{equation*}
1\,\ll \,n\,\ll \,\frac{Q}{\Lambda }
\end{equation*}
or equivalently 
\begin{equation}
\Lambda \,Q\,\ll \,m^{2}\,\ll \,Q^{2},
\end{equation}
the QCD function, the shape function and the coefficient function are all
reliably computed in perturbation theory. Actually, in this case there is no
need to introduce the shape function and the resummed QCD form factor is all
is needed: its splitting in a shape function and a coefficient function is
irrelevant;

\item  In the ``more exclusive'' region 
\begin{equation*}
\frac{Q}{\Lambda }\,\sim \,n\,\ll \,\frac{Q\,\mu }{\Lambda ^{2}}
\end{equation*}
or equivalently 
\begin{equation}
\Lambda \,Q\,\sim \,m^{2}\,\gg \,\frac{Q\,\Lambda ^{2}}{\mu },
\end{equation}
the QCD function and the shape function become non-perturbative, while the
coefficient function is still perturbatively computable. The idea is that
one replaces the perturbative evaluation of the shape function with a
non-perturbative one and \textit{defines} the QCD distribution by means of
relation (\ref{fondamento});

\item  If one considers larger moments or, equivalently, smaller jet masses, 
\begin{equation}
n\,\gtrsim \,\frac{Q\,\mu }{\Lambda ^{2}}\qquad \mathrm{or}\qquad
m^{2}\,\lesssim \,\frac{Q\,\Lambda ^{2}}{\mu },
\end{equation}
\textit{also} the coefficient function becomes non-perturbative and the
shape function loses most of its meaning.
\end{enumerate}

The coefficient function $C_{n}$ is plotted in fig.\thinspace 1 together
with the QCD function $\mathbf{f}_{n}.$ For small values of $n$ the two
curves are very close to each other because typical transverse momenta are
large and the lower cut-off $\mu $ in $C_{n}$ is ineffective. In the figure
we have taken $Q=5.2\,$GeV, $\alpha _{S}=0.24,\,\mu =2\,$GeV, $\Lambda
=0.3\, $GeV and $n_{f}=3.$ For these value of the parameters, $\mathbf{f}%
_{n} $ becomes singular for $n>17$, while $C_{n}$ becomes singular for $%
n>115.$ The coefficient function monotonically decreases with $n,$ starting
from $C_{n}=1 $ at $n=1$ and going down to $C_{n}\simeq 0.68$ for $n=17.$
This decay is produced by gluons of energy between the cutoff $\mu $\ and
the hard scale $Q.$ The function $\mathbf{f}_{n}$ also monotonically
decreases with $n$ starting again from $\mathbf{f}_{n}=1$ at $n=1$ and
reaching $\mathbf{f}_{n}\simeq 0.33$ for $n=17.$ This faster decay is
produced by gluons with any energy between the soft scale $|k_{+}|$ and $Q.$

The result (\ref{lei}) applies also to the coefficient function of the shape
function defined in lattice regularization\thinspace \cite{lattice},\ after
the identification is done 
\begin{equation}
\frac{1}{a}=c\,\mu ,
\end{equation}
where $c$ is a constant of order $1$ and $a$ is the lattice spacing. The
precise value of $c$ is determined only in next-to-leading approximation.
One has to perform a full order $\alpha _{S}$ computation of the shape
function in lattice regularization and equate it to the expansion to order $%
\alpha _{S}$ of the next-to-leading analogue of eq.\thinspace (\ref
{ancoralei}).

\section{Conclusions}

In general, the QCD semi-inclusive form factor $\mathbf{f}\left( z\right) $
factorizes and resums large infrared double-logarithms which occur in the
threshold region 
\begin{equation}
m^{2}\,\ll \,Q^{2}.
\end{equation}
These terms are related to the emission of soft gluons with transverse
momenta ranging from the infrared scale $|k_{+}|=m^{2}/Q$ up to the hard
scale $Q.$ As long as one keeps $|k_{+}|\gg \Lambda ,$ dynamics is
controlled by perturbation theory. A phenomenologically relevant region is,
however, $|k_{+}|\sim \Lambda ,$ in which substantial non-perturbative
effects are encountered. We factorize them by means of the shape function.
An additional, unphysical, scale $\mu $ is introduced in the problem,
intermediate between the hard and the soft scale: 
\begin{equation}
|k_{+}|\,\ll \,\mu \,\ll \,Q.
\end{equation}
This means a splitting of the double logarithm in the QCD cumulative
distribution of the form 
\begin{eqnarray}
\mathbf{F} &=&1-\frac{A_{1}\alpha _{S}}{2}\log ^{2}\frac{Q^{2}}{m^{2}}=1-%
\frac{A_{1}\alpha _{S}}{2}\log ^{2}\frac{Q}{|k_{+}|}  \notag \\
&=&\left( 1-\frac{A_{1}\alpha _{S}}{2}\log ^{2}\frac{Q}{\mu }-A_{1}\alpha
_{S}\log \frac{Q}{\mu }\log \frac{\mu }{|k_{+}|}\right) \left( 1-\frac{%
A_{1}\alpha _{S}}{2}\log ^{2}\frac{\mu }{|k_{+}|}\right) .  \label{ba}
\end{eqnarray}
The scale $\mu $ has the role of an IR cut-off for the coefficient function
and of an UV cut-off for the shape function. The first parenthesis in (\ref
{ba}) contains the coefficient function, which is short-distance-dominated
and reliably computed in PT, while the second parenthesis contains a
perturbative evaluation of the shape function. The latter computation is
``unreliable'', so it is thrown away and replaced by a non-perturbative one.
The factorization scale $\mu $ then separates what we compute in PT (above $%
\mu $) from what we give up to compute in PT (below $\mu $). We may say that
the coefficient function of the shape function is the ``upper part'' of a
semi-inclusive QCD form factor.

We have shown that the transverse gluon momenta entering the coefficient
function have a lower bound given by\footnote{%
The appearance of a single logarithm of \ the soft scale $|k_{+}|$ in the
coefficient function, eq.\thinspace (\ref{ba}), should not be erroneously
interpreted as the signal of a non-perturbative effect in $C$.} 
\begin{equation}
l_{\perp }\,\gtrsim \,\sqrt{|k_{+}|\,\mu }\,\sim \sqrt{\Lambda \,\mu }\,\gg
\,\Lambda ,
\end{equation}
where eq.\thinspace (\ref{cinshape}) has been used. As a consequence,
non-perturbative corrections to the factorized form (\ref{fondamento2}) or (%
\ref{fondamento})\ are expected to be of the size 
\begin{equation}
\frac{\Lambda }{\mu },
\end{equation}
involving one inverse power of the factorization scale $\mu $. This
situation is to be contrasted with single logarithmic problems, where
corrections to factorization have typically a size 
\begin{equation}
\frac{\Lambda ^{2}}{\mu ^{2}},
\end{equation}
involving two inverse powers of the factorization scale. We may say that
factorization in this double-logarithmic problem is still consistent, but it
is presumably ``less lucky'' than in single-logarithmic cases.

We conjecture that a similar factorization --- into a perturbatively
calculable coefficient function and a non-perturbative component --- can be
done for other semi-inclusive distributions, such as shape variables for
very small values of the resolution parameters. An example is the thrust
distribution in $e^{+}e^{-}$ annihilations \cite{thrust}\ for values of the
jet masses in the region (\ref{cinshape}), for which 
\begin{equation}
1-T\,\sim \,\frac{\Lambda }{Q}.
\end{equation}
In other words, we believe that the idea of the shape function can be
generalized to describe many semi-inclusive distributions in the region (\ref
{cinshape}).

To conclude, we have presented a leading evalutation of the resummed
coefficient function of the shape function and we have proved that the
latter is short-distance-dominated.

\begin{center}
Acknowledgements
\end{center}

I wish to thank S. Catani and L. Trentadue for suggestions. I acknowledge
also discussions with G. Buchalla, M. Ciuchini, P. Gambino, M. Luke and T.
Mannel.

\end{document}